# Temperature Dependent Behavior of Thermal Conductivity of Sub-5 nm Ir film: Defect-electron Scattering Quantified by Residual Thermal Resistivity

Zhe Cheng,[1] Zaoli Xu,[1] Shen Xu,[1] and Xinwei Wang[1,2,a)]

[1]*Department of Mechanical Engineering, 2010 Black Engineering Building,*

*Iowa State University, Ames, Iowa, 50011, USA*

[2]*School of Urban Development and Environmental Engineering, Shanghai Second Polytechnic University, Shanghai 201209, P.R. China*

By studying the temperature-dependent behavior (300 K down to 43 K) of electron thermal conductivity ($\kappa$) in a 3.2 nm-thin Ir film, we quantify the extremely confined defect-electron scatterings and isolate the intrinsic phonon-electron scattering that is shared by the bulk Ir. At low temperatures below 50 K, $\kappa$ of the film has almost two orders of magnitude reduction from that of bulk Ir. The film has $\partial\kappa/\partial T>0$ while the bulk Ir has $\partial\kappa/\partial T<0$. We introduce a unified thermal resistivity ($\Theta=T/\kappa$) to interpret these completely different $\kappa\sim T$ relations. It is found that the film and the bulk Ir share a very similar $\Theta\sim T$ trend while they have a different residual part ($\Theta_0$) at 0 K limit: $\Theta_0\sim0$ for the bulk Ir, and $\Theta_0=5.5$ m·K$^2$/W for the film. The Ir film and the bulk Ir have very close $\partial\Theta/\partial T$ (75 to 290 K): $6.33\times10^{-3}$ mK/W for the film and $7.62\times10^{-3}$ mK/W for the bulk Ir. This strongly confirms the similar phonon-electron scattering in them. Therefore, the residual thermal resistivity provides an unprecedented way to quantitatively evaluating defect-electron scattering ($\Theta_0$) in heat conduction. Moreover, the interfacial thermal conductance across the grain boundaries is found larger than that of Al/Cu interface, and its value is proportional to temperature, largely due to the electron's specific heat. A unified interfacial thermal conductance is also defined and firmly proves this relation. Additionally, the electron reflection coefficient is found to be large (88%) and almost temperature independent.

[a] Corresponding author. Email: xwang3@iastate.edu, Tel: 515-294-2085, Fax: 515-294-3261

## I. INTRODUCTION

Metallic ultra-thin films are widely used as interconnects in the microelectronic industry and play an important role in related thermal design in micro/nanoscale devices and systems.[1] The performance of these applications is significantly affected by the energy transport and dissipation in these metallic films. These films are composed of nanocrystals and electron dominates in their thermal transport. When the film dimension is either comparable to or less than the electron mean free path, the grain boundaries and surfaces scatter electrons strongly.[2,3] As a result, the thermal properties of metallic ultra-thin films behavior quite differently from their bulk counterparts, especially at low temperatures when the phonon-electron scattering diminishes gradually. The thinner the film thickness is, the stronger grain boundary and surface scatterings are. Subsequently, the thermal behavior difference between the film and their bulk counterpart becomes larger. Data for these extremely confined domains will be in high demand in the future applications. However, due to the difficulties in sample preparation and accurate in-plane thermal conductivity characterization of nanometer-thick metallic films at low temperatures, few experimental reports are available. Yoneoka *et al.* measured thermal conductivity of platinum films with a thickness of 7.3, 9.8, and 12.1 nm from 320 K to 50 K.[4] Zhang and co-workers investigated the thermal transport in 53 nm and 76 nm thick Au nanofilms from 300 K to 3 K and 48 nm thick platinum nanofilms from 300 K to 60 K.[5,6] It is noticeable that the thinnest metallic film whose temperature dependent thermal conductivity has been measured so far is the 7.3 nm platinum film studied by Yoneoka and coworkers. For extremely thin films (sub-5 nm thick), the temperature dependent nature of thermal conductivity has not been studied before, even though such work is crucial for the detailed understanding of electron thermal transport with extremely strong defect scatterings at low temperatures. Therefore, it is of great importance to extending the thickness limit and an in-depth study of energy dissipation and transport in the sub-5 nm regime is overdue.

In this work, a robust and accurate technique [7,8] named transient electro-thermal (TET) technique, developed in our lab is used to characterize the thermal transport properties in ultra-thin metallic films. The thermal conductivities of nanocrystalline Iridium (Ir) films having an average thickness of 3.2 nm are studied from 300 K to 43 K. The temperature dependent thermal conductivity is investigated and compared with that of bulk Ir to reveal the extremely strong structural scattering effect. A unified thermal resistivity is introduced to interpret the completely different thermal conductivity variation trends against temperature of the film and the bulk Ir.

## II. SAMPLE STRUCTURE

The ultra-thin Ir films studied in this work cannot support themselves due to its very fine thickness. Therefore, milkweed floss is selected as the substrate to support the ultra-thin Ir films for reasons given later. The milkweed floss is collected from a dry milkweed seed pod grown in Ames, Iowa, USA. The milkweed seeds and floss are shown in Figure 1(a). Figure 1(b) depicts the SEM image of a single milkweed fiber suspended across two electrodes. The two ends of the fiber are long enough to avoid being embedded in the silver paste. This ensures that the silver paste will not enter the hollow part of the fiber. The inset shows the smooth floss surface. Figure 1(c) depicts the SEM image of the milkweed fiber cross section. The definition of the maximum Ir film thickness $\delta_{\max}$, diameter $d$ and cell wall thickness $\delta_{floss}$ are shown in Figure 1(d). The average thickness of Ir films is $\delta_{ave} = 2\delta_{\max}/\pi$. During the Ir film deposition process using argon-ion discharge sputtering, the Ir atoms will deposit on the floss like snow precipitation. This makes the Ir film have the largest thickness on the top, and the least one on the side [as shown in Figure 1(d)]. The measured properties are the effective properties of the overall films whose thickness ranges from 0 to the largest thickness. Afterwards, if not specially mentioned, the thickness will be the mean average thickness. In this work, the Ir films on milkweed fibers are coated using a sputtering machine (Quorum Q150T S). The thicknesses ( $\delta_{\max}$ ) of the deposited Ir films are monitored using a quartz crystal microbalance. The accuracy of the thickness measurement is verified by an atomic force microscope.

It should be pointed out that the floss surface is not atomic-level smooth although the Ir sputtering machine can deposit very fine grains on the floss surface. Also the sputtered layer cannot reach atomic level uniformity. Therefore, the average thickness refereed in this work represents an average value: a value that is obtained by the deposition mass divided by the projected area in the deposition direction. Still the film shows high-level thickness uniformity as shown and discussed later in Figure 6(b), which will be discussed later. From that figure, it is clear the Ir film (32 nm average thickness in TEM study) is continuous along the surface of the floss surface, and it shows nm-scale surface smoothness.

Here we choose milkweed floss as the substrate material due to several reasons. First, the milkweed floss is a unique natural cellulose fiber that has a low density due to the presence of a completely hollow center.[9-11] No other known natural cellulose fiber has such an overall low density.[9] Consequently it will have a very low overall thermal conductivity. This will provide a great advantage for studying the Ir film on it because the overall thermal diffusivity would have a great increase even when a very thin Ir film is deposited on it. Second, the fiber surface is smooth and its diameter is very uniform and well defined, as shown in the inset of Figure 1(b). This ensures accurate control and measurement of the metallic film's geometry. Milkweed floss is a single-cell fiber,[9] so our experiment can provide fundamental information about the energy transport capacity along single plant cell wall as the byproduct. Furthermore, milkweed floss has been used or reported as textiles and filling material.[11-14] Plant cell fibers composed of cellulose and lignin also could be an excellent platform for flexible electronics. Therefore, it is of great interest to investigate the heat conduction in Ir films grown on it, as well as its own thermal properties.

As shown in Figure 1(c), the milkweed fiber is hollow. Under a scanning electron microscope (SEM), the average milkweed wall thickness is determined as 614 nm. In this work, four sets of experiments are conducted from room temperature down to 10 K. 10 K is the lowest temperature the sample could stay. When the temperature is lower than 43 K, the electrical resistance does not change with temperature linearly, and also has very weak temperature dependence. Therefore, the TET technique cannot be used to characterize the thermal diffusivity accurately. First, after the milkweed fiber is coated with the first Ir layer with an average thickness of 9.6 nm, the effective thermal diffusivity is measured from room temperature to 43 K. Then the temperature is allowed to rise slowly to room temperature. We have confirmed that the electrical resistance of the sample at room temperature remains unchanged after the sample experiences the extremely low temperature environment. This firmly concludes that the structure of the milkweed and Ir film on it is unchanged in our thermal characterization from room temperature to 10 K. After the first round of measurement is done, a second layer of Ir with an average thickness of 3.2 nm (whose $\delta_{max}$ is 5 nm) is coated. Subsequently, the measurement is repeated from room temperature to 10 K. Then again the temperature goes back to room temperature slowly. These measurement processes are repeated four times and the third and fourth Ir layers are the same as the second one. During these processes, the structure of milkweed and Ir films are not affected by the low temperature. This is critical to ensure the properties of the four ultra-thin films are the same.

## III. THERMAL TRANSPORT CHARACTERIZATION

### A. Differential methodology to characterize the thermal transport in ultra-thin Ir films

A robust and advanced differential technology,[7,8] has been developed in our lab to characterize the thermal properties of ultra-thin metallic films. The measured film thickness can reach sub-5 nm, even sub-nm while other technologies cannot achieve this level. In this work, a milkweed fiber is suspended across two electrodes as the supporting material for the ultra-thin metallic films as shown in Figure 2(b).

For thermal characterization of a one-dimensional material by using the TET technique, the material has to be electrically conductive. Therefore, the milkweed fiber is first coated with a Ir film of thickness $\delta_1$ (the first layer) and the effective thermal diffusivity of the milkweed fiber-metallic film system in the axial direction is measured as $\alpha_{eff,1}$. Then the same sample is coated with a second Ir layer of thickness $\delta_2$, and the whole sample's thermal diffusivity is measured again as $\alpha_{eff,2}$. The thermal diffusivity increment induced by the second Ir layer is $\Delta\alpha_{eff} = \alpha_{eff,1} - \alpha_{eff,2}$. This thermal diffusivity differential is directly related to the Lorenz number of the second Ir layer of thickness $\delta_2$, and other parameters of the sample, like the milkweed fiber's geometry and thermal properties. Theoretically, to measure the electrical and thermal conductivities, and the Lorenz number of the second Ir layer of thickness $\delta_2$, only one second layer ($\delta_2$ thickness) needs to be coated. To improve the measurement accuracy and significantly suppress experimental uncertainty, we repeatedly deposit Ir layers of thickness $\delta_2$ and measure the corresponding thermal diffusivity $\alpha_{eff,n}$.

After that, the thermal conductivity of a single Ir layer of thickness $\delta_2$ is determined based on the increment of thermal diffusivity ($\Delta\alpha_{eff}$). Here, both $\delta_1$ and $\delta_2$ refer to the maximum thickness of the Ir films. The first Ir layer ($\delta_1$ thickness) is used to make the sample electrically conductive. So the thickness of this layer can be the same or different from $\delta_2$. In this work, $\delta_1$ is chosen to be 15 nm, which is thick enough to obtain a stable electrical resistance of the sample. $\delta_2$ is 5 nm and three layers of Ir films with thickness of $\delta_2$ are deposited layer by layer on the first layer. It is physically reasonable that each deposited Ir layer ($\delta_2$ thickness) has the same thermal properties because they have the same thickness and are deposited under the exactly same conditions. This assumption is fully checked and verified by the experimental results and discussed later. Details of the theory and experimental process for this differential technology are given in below.

The measured thermal diffusivity ($\alpha_{eff}$) is an effective value combining both effects of the milkweed fiber and Ir coatings.

$$\alpha_{eff} = \frac{A_m\kappa_m + A_1\kappa_1 + nA_2\kappa_2}{A_e(\rho c_p)_e}, \tag{1}$$

where $n$ is the number of $\delta_2$ thick layers. $A_e$, $A_m$, $A_1$, and $A_2$ are the cross-sectional area of the coated fiber, bare fiber (including the hollow region), the first Ir layer, and an individual $\delta_2$ thickness Ir layer. The thin Ir layer has negligible contribution to the overall cross-sectional area of the sample, so we have $A_e$=$A_m$. Moreover, the contribution of ultra-thin Ir films to volumetric specific heat is negligible (~1%), so we take the volumetric specific heat to be unchanged ($(\rho c_p)_e$=$(\rho c_p)_m$). $\alpha_{eff}$ increases with the number of film layers and the slope is $A_2\kappa_2/A_e(\rho c_p)_e$. $A_2/A_e(\rho c_p)_e$ is known already so the thermal conductivity ($\kappa_2$) of the 3.2 nm-thick Ir film can be determined.

## B. Thermal characterization of Ir-coated floss

The TET technique [15,16] developed in our laboratory is used to measure the effective thermal diffusivity ($\alpha_{eff}$) of the Ir-covered milkweed fiber. A schematic of the TET technology is presented in Figure 2(a). The to-be-measured sample is suspended across two aluminum electrodes, and placed in the vacuum chamber of a cryogenic system (CCS-450, JANIS). To eliminate heat convection in the measurement, a liquid nitrogen cold-trapped mechanical vacuum pump is used to reach a vacuum level of 0.4 mTorr. During thermal characterization, a step DC current is fed through the sample to generate electric heat that induces a temperature rise of the sample. The temperature rise of the sample will induce an electrical resistance change, which leads to an overall voltage change. Therefore, the voltage change of the sample can be used to monitor its temperature evolution, and determine the thermal diffusivity of the sample. Details of the experimental process and data reduction are given in below.

During TET thermal characterization, the average temperature along the sample can be expressed as:

$$\bar{T} = T_0 + \frac{q_0 L^2}{12}\frac{48}{\pi^4}\sum_{m=1}^{\infty}\frac{1-(-1)^m}{m^2}\frac{1-\exp\left[-\left(m^2-f\right)\pi^2(\alpha t/L^2)\right]}{(m^2-f)}. \quad (2)$$

As time goes to infinity, the temperature distribution along the sample will reach a steady state. The average temperature of the sample in the final steady state is:

$$T(t \to \infty) = T_0 + \frac{q_0 L^2}{12k}. \tag{3}$$

More details for the above equation's derivation are provided in references.[15,17] With an effective thermal diffusivity $\alpha_{eff}=\alpha\cdot(1-f)$, here $f$ is defined as $-8\varepsilon_r \sigma T_0^3 L^2 / d\pi^2 k$ (the radiation effect), the normalized average temperature rise $T^*$ is:

$$T^* \cong \frac{48}{\pi^4} \sum_{m=1}^{\infty} \frac{1-(-1)^m}{m^2} \frac{1-\exp[-m^2\pi^2\alpha_{eff} t / L^2]}{m^2}. \tag{4}$$

The measured voltage change is inherently proportional to the temperature change of the sample. The normalized temperature rise $T^*$ is calculated from experiment as $T^* = (V_{sample} - V_0)/(V_1 - V_0)$, where $V_0$ and $V_1$ are the initial and final voltages across the sample. In our work, after $T^*$ is obtained, different trial values of $\alpha_{eff}$ are used to calculate the theoretical $T^*$ using Eq. (4) and fit with the experimental result. The value giving the best fit of $T^*$ is taken as the effective thermal diffusivity of the sample.

Here we take the sample at room temperature as an example to demonstrate how the effective thermal diffusivity is characterized. The length and diameter of this sample is 981 μm and 20.53 μm respectively. The sample is coated with the first Ir layer ($\delta_{1,ave}$=9.6 nm). The electrical resistances before and after applying a step current are 615.99 Ω and 623.15 Ω. The electrical current used in the experiment is 156 μA. This gives a voltage change at about 1% due to self-joule heating. Figure 2(c) shows the transient voltage change of raw experimental data. The normalized temperature rise and the fitting result are shown in Figure 2(d). The effective thermal diffusivity is determined as $1.03\times10^{-6}$ $m^2\cdot s^{-1}$, which includes the effect of radiation and parasitic conduction. We vary the trial values of $\alpha$ to determine the fitting uncertainty as shown in Figure 2(d). When the trial value is changed by 10%, the theoretical results deviate from the experimental results evidently. It is evident that the experimental data falls within a range of ±10% of the theoretical fitting.

## IV. THERMAL TRANSPORT IN AN INDIVIDUAL 3.2 NM-THICK IR FILM

### A. Effective thermal diffusivity increments by the Ir films

The effective thermal diffusivity of the sample is characterized with the TET technique from room temperature down to 10 K. When the temperature is lower than 43 K, the electrical resistance does not change with temperature linearly, and also has very weak temperature dependence. Therefore, the TET technique cannot be used to characterize the thermal diffusivity accurately. The measurement results are shown in Figure 3. The effective thermal diffusivity increases with decreasing temperature. The lower the temperature is, the faster the thermal diffusivity rises.

As we can see from Figure 3, the effective thermal diffusivity increases by the same amount when each average 3.2 nm-thick Ir film is added on the sample. The solid curves represent the trends of the effective thermal diffusivity change with temperature. Every time an average 3.2 nm-thick Ir film is added, the effective thermal diffusivity increment is denoted as $\Delta\alpha_{eff,1}$. The inset in Figure 3 shows the change of thermal diffusivity at room temperature against the number of average 3.2 nm-thick Ir films. An excellent linear trend is observed. This strongly proves that each layer has the same thermophysical property. The effective thermal diffusivity increment induced by each Ir layer at low temperatures bears a little more noise/uncertainty. Therefore, we use the effective thermal diffusivity increment ($\Delta\alpha_{eff}$) between the fourth layer and the first layer case to determine the thermal transport properties of the Ir film. This data treatment maximizes the thermal diffusivity difference and efficiently suppresses measurement uncertainty. As we can see from Figure 3, $\Delta\alpha_{eff}$ shows weak temperature dependence and changes linearly with temperature. Also the uncertainty in the data becomes small, which is more tolerable. A linear fitting is used to smooth the effective thermal diffusivity difference $\Delta\alpha_{eff}$ and the result will be used for thermal conductivity determination of the Ir film.

## B. Thermal conductivity and unified thermal resistivity of the Ir film

Based on $\Delta\alpha_{eff}$ induced by the three 3.2 nm-thick Ir films, we could find the thermal conductivity $\kappa$ of an individual Ir film as $\kappa = (\rho c_p)_m \cdot (A_m/A_2) \cdot (\Delta\alpha_{eff}/3)$. In this equation, $A_m/A_2$ is the cross-sectional area ratio of the sample to the 3.2 nm-thick Ir film. The effective thermal diffusivity increment induced by each 3.2 nm-thick Ir film is $\Delta\alpha_{eff}/3$ because each 3.2 nm-thick Ir film has the same thermal conductivity, which is verified by experimental results shown in Figure 3. $(\rho c_p)_m$ is the effective volumetric specific heat of the sample. To determine the thermal conductivity of an individual Ir layer, the effective volumetric specific heat $(\rho c_p)_m$ of the sample is needed. So we must determine this property in advance. Details on how this property is determined are given in below.

During TET characterization, the average temperature rise is $\Delta T = q_0 L^2 / 12k_{eff}$ according to Eq. (3), here $q_0 = 4I^2 R / \pi d^2 L$ is the heat generation per unit volume. The temperature rise during our TET characterization can be obtained from the electrical resistance change ( $\Delta R$ ) as $\Delta T = \Delta R/(\eta R_0)$ . $\eta$ is the temperature coefficient of resistance. Then we can obtain the effective thermal conductivity $k_{eff} = q_0 L^2 / 12\Delta T$ . As the effective thermal diffusivity has been determined for each round of experiment (Figure 3), the volumetric specific heat $(\rho c_p)_m = \kappa_{eff} / \alpha_{eff}$ can be obtained. The volumetric specific heat of the milkweed fiber is determined four times because the contribution of the ultrathin metallic films to the total volumetric specific heat is negligible (the maximum average contribution of Ir is ~1%).

The specific heat shown in Figure 4 shows a good linear relation with temperature, so a linear fitting is used to smooth the data for later use. The volumetric specific heat of milkweed fiber decreases with decreasing temperature. This is because the short wave phonons are frozen out and only the long wave phonons are excited to contribute to the specific heat. In Figure 4, the volumetric specific heat of milkweed fiber is compared with that of the microcrystalline cellulose in literature.[18] It can be seen that the two lines overlap using different vertical coordinates, which means the trends of volumetric specific heat against temperature are the same for the milkweed fiber and the microcrystalline cellulose. Using the volumetric specific heat of microcrystalline cellulose as the reference, then we can obtain the volumetric ratio of the cell wall as 12.85%. Subsequently, the cell wall thickness is determined as 660 nm, which is only 7% larger than our SEM measurement result (614 nm). The measured thickness (614 nm) is smaller than the expected thickness (660 nm) based on the assumption that the thermal property of milkweed cell wall is similar to that of the microcrystalline cellulose. This means the real volumetric specific heat of the milkweed cell wall is larger than that of the microcrystalline cellulose. This can be explained by several reasons. First, the contribution of metallic film gives some contribution to this difference. Furthermore, the specific heat of the amorphous state is usually larger than that of the crystalline state. Finally, the cell wall contains other materials apart from cellulose, like lignin and hemicellulose. The specific heats of these materials also affect the overall specific heat.

The inset in the bottom right corner of Figure 4 shows the effective thermal conductivity for the four cases. The real thermal conductivity of the milkweed fiber (including the effect of the hollow center) can be obtained by subtracting the effects of the Ir films and radiation. The intrinsic thermal conductivity of the milkweed fiber can be determined after the volumetric ratio of the milkweed cell wall is known. Details are described below and the results are shown in the inset in the upper left corner of Figure 4.

To obtain the real and intrinsic thermal conductivity of the milkweed floss, we need to subtract the effect of radiation. Experiment on a long sample is conducted to determine the surface emissivity $\varepsilon_r$ based on $k_{real} = \alpha_{eff,n}\rho c_p - 16\varepsilon_r \sigma_{SB} T_0^3 L_n^2 / (d_n \pi^2) - 4L_{Lorenz} T_n L_n / (R_n \pi d_n^2)$ . Subscript $n$ takes 1, 2, referring to the long sample and the sample used above, respectively. $\sigma_{SB}$ is the Stefan-Boltzmann constant. We use the volumetric specific heat of the sample used above [$2.33\times10^5$ J/(K m$^3$)] and the Lorenz number [$2.29\times10^{-8}$ W·Ω·K$^{-2}$] to calculate the emissivity. The length and diameter of the long sample are 2366 μm and 23.99 μm respectively. The effective thermal diffusivity of the long sample and the sample used above are $1.71\times10^{-6}$ m$^2$ s$^{-1}$ and $1.00\times10^{-6}$ m$^2$ s$^{-1}$ respectively. Only the real thermal conductivity of the milkweed fiber $k_{real}$ and surface emissivity $\varepsilon_r$ are unknown in the two equations above and the surface emissivity is determined as 0.40.

The real thermal conductivity of the milkweed fiber can be determined after subtracting the effect of radiation and parasitic conduction of the metallic film as $\kappa_{real} = \kappa_{eff} - 16\varepsilon_r \sigma_{SB} T_0^3 L^2 / (d\pi^2) - 4L_{Lorenz} TL / (R\pi d^2)$ . The real thermal conductivity of the milkweed fiber is obtained and depicted in the inset in the upper left corner of Figure 4. Because we know the thickness of the cell wall through SEM measurement, the volumetric ratio is obtained as 14.78%. The intrinsic thermal conductivity of the milkweed cell wall is also determined by dividing $k_{real}$ by 14.78% and shown in the inset in the upper left corner of Figure 4 using the right coordinate axis.

The thermal conductivity of milkweed fiber decreases with decreasing temperature. The thermal conductivity shows a similar trend with the volumetric specific heat, namely changing linearly with temperature. Theoretically, we can determine the thermal conductivity of an individual Ir film by examining the thermal conductivity increment by the addition of each Ir film. However, as shown in the inset in the bottom right corner of Figure 4, the thermal conductivity data barely reveals good enough data to calculate the increment. The effective thermal conductivity data carries much more uncertainties than thermal diffusivity. This is because the thermal conductivity evaluation relies on more data, like electrical resistance temperature coefficient, resistance rise in experiment, and electrical heating level. So we do not use the directly measured effective thermal conductivity to evaluate the thermal transport in the Ir film.

Based on the volumetric specific heat (Figure 4) and $\Delta\alpha_{eff}$ (Figure 3), the thermal conductivity ($\kappa$) of an individual 3.2 nm-thick Ir film is calculated and shown in the left inset of Figure 5. The orders of magnitude reduction of the film's thermal conductivity in comparison with that of the bulk Ir is also obtained and shown in the right inset of Figure 5. It can be seen that the thermal conductivity of the 3.2 nm-thick Ir film is significantly reduced from the bulk value. When temperature goes below 50 K, the thermal conductivity reduction reaches an extremely high level: close to two orders of magnitude. This is due to strong grain boundary scattering, which limits the electron mean free path significantly. For the bulk Ir, the short wave phonons are frozen out. Only long wave phonons contribute to the phonon-electron scatterings at low temperature. The decrease of scattering sources results in the increase of electron mean free path and subsequently thermal conductivity. The film's thermal conductivity decreases with decreasing temperature. This trend is completely opposite to that of the bulk Ir. This kind of phenomenon also has been observed in gold and platinum nanofilms,[4,19,20] nickel nanowire[21] and alloys[22]. The reduced thermal conductivity was attributed to the increased scatterings of heat carriers from structural imperfection and the contribution of phonon thermal conductivity.[4,21] Here we will provide an explanation of the abnormal temperature dependent thermal conductivity of these metallic nanostructures.

The thermal conductivity of electrons can be expressed as $\kappa = C_V v_F^2 \tau / 3$. Here $C_v$ is the volumetric electron heat capacity; $v_F$ is the Fermi velocity and $\tau$ is the relaxation time. Besides the electron's relaxation time, the thermal conductivity is strongly and directly affected by temperature. This effect stems from the electron heat capacity in the thermal conductivity relation. The heat capacity is $C = \gamma T$ where $\gamma$ is 3.1 mJ·mol$^{-1}$·K$^{-2}$ for Ir when the temperature is not too high.[23] The temperature in the thermal conductivity's expression overshadows the physics behind the variation of $\kappa$ against $T$. The traditional thermal resistivity is defined as $W=\kappa^{-1}=3/\gamma T v_F^2 \tau$ . Instead of directly looking at *W*, we define a unified thermal resistivity: $\Theta = W \times T$ . It is clear this unified thermal resistivity is solely related to the electron relaxation time ($\tau$). This unified thermal resistivity plays the same critical role as the electrical resistivity in reflecting the electron scattering in metals.

We plot out the unified thermal resistivity variation against temperature in comparison with the bulk's values,[24,25] as depicted in Figure 5. "3.2 nm" is for the Θ of 3.2 nm-thick Ir film calculated from $\Delta\alpha_{eff}$ shown in Figure 3. One striking phenomenon is that the unified thermal resistivity follows a very similar trend to the behavior of electrical resistivity. When temperature is extended to 0, Θ of bulk Ir is almost 0 with a negligible residual value. For the Ir film, it has a residual value of about 5.5 $mK^2/W$ [$\Theta_0$]. This value makes the dominant contribution to the overall Θ. At room temperature, the overall Θ is only about 7 $mK^2/W$. Also the unified thermal resistivity of the 3.2 nm thick Ir and the bulk Ir share the similar trend against temperature, although the one of 3.2 nm thick Ir has a smaller slope. When the temperature approaches zero, both reach a constant value (residual resistivity) while the Ir film has a much larger residual value. This trend similarity is totally different from that of the thermal conductivity comparison in the left inset of Figure 5. In the left inset, no observable conclusion can be made about the comparison since the thermal conductivity of the Ir film and the bulk Ir shows totally different absolute values and a totally different trend of variation against temperature. Therefore, the unified thermal resistivity Θ is a critical property to reflect the electron scattering that determines thermal transport. The comparison with that of its bulk counterpart provides a great way to evaluating the effect of structural defects on electron thermal transport.

Now we can explain the completely different trend of $\kappa$ variation against $T$ for the Ir film compared with bulk Ir. The left inset shows that the thermal conductivity of bulk Ir rises sharply at low temperature. That is because the residual part of bulk Ir ($\Theta_0$) is close to zero at low temperatures. Unlike that of bulk Ir, $\Theta_0$ of the 3.2 nm-thick Ir film (about 5.5 mK$^2$/W) is much larger than the temperature dependent part $\Theta_{el\text{-}ph}$ (1.57 mK$^2$/W) at room temperature. Moreover, the effect of the temperature dependent part diminishes with decreasing temperature. This means the effect of $\Theta_0$ increases with decreasing temperature. All of these factors contribute to the decreased thermal conductivity of the Ir film when temperature decreases.

## C. Characteristic structure size for electron scattering

Like the electrical resistivity, the classical thermal resistivity is also composed of two parts: $W = W_0 + W_{el-ph} = 3(\tau_0^{-1} + \tau_{el-ph}^{-1}) \big/ (\gamma T v_F^2)$. Here, subscripts "0" and "*el-ph*" represent the thermal resistivity induced by the structural imperfections and by phonon scattering respectively. According to Matthiessen's rule and relaxation time approximation of scatterings, the unified thermal resistivity can be expressed as $\Theta = \Theta_0 + \Theta_{el-ph} = 3(\tau_0^{-1} + \tau_{el-ph}^{-1}) \big/ (\gamma v_F^2)$. So $\Theta$ is composed of two parts: the residual part $\Theta_0$ that is temperature independent, and the temperature dependent part $\Theta_{el\text{-}ph}$. Similar to electrical resistivity, we define the slope of $\Theta$ variation against temperature as the temperature coefficient of thermal resistivity (TCTR). The TCTR of the 3.2 nm thick Ir film ($6.33\times10^{-3}$ mK/W from 290 K to 75 K) is a little smaller than, but still close to that of the bulk material ($7.62\times10^{-3}$ mK/W from 290 K to 75 K). This strongly proves they share the similar phonon-electron scattering. The unified thermal resistivity goes down with decreasing temperature due to the reduction of phonon density. This behavior is very similar to that of electrical resistivity (to be detailed in future publication).

The residual part of the 3.2 nm thick Ir film (about 5.5 $mK^2/W$) is much larger than that of the bulk material ($1.4\times10^{-3}$ $mK^2/W$). The Fermi energy of Ir is 0.761 R*y*.[26] The Fermi velocity can be determined as $1.91\times10^6$ m/s [$v_F = (2E_F/m_e)^{0.5}$]. Then we can obtain the value of $\tau_0$ is $3.8\times10^{-16}$ s. Finally the mean free path ($l_0$) at low temperatures is determined as 0.73 nm ($l_0 = \tau_0 v_F$). At low temperatures, the effect of phonon-electron scattering diminishes. The structure scatterings, like grain boundary scattering, surface scattering and point defect scattering, dominate the electron transport. Therefore, the calculated $l_0$ gives a characteristic structure size that scatters electrons during heat conduction. The crystalline size of the thin films is estimated to be about 8 nm according to the XRD results. The characterization details will be described later. This size is much larger than the film thickness, proving that the film has columnar structure in the vertical direction. The size given by XRD represents the characteristic size of the columns in the lateral (in-plane) direction of the film. This is also the electron heat conduction direction studied in this work.

The above revealed nanocrystalline structure of the Ir film is confirmed by high-resolution transmission electron microscopy study. X-ray diffraction (XRD) is also used to characterize the structure of milkweed fibers. The XRD system (Siemens D 500 diffractometer) is equipped with a copper tube that was operated at 40 kV and 30 mA. Because one milkweed fiber is too small compared with the XRD spot size, we use a bunch of milkweed fibers and align them parallel to each other. These fibers are suspended and scanned by XRD. They are confirmed amorphous. To obtain the structure information of the Ir film, a layer of 3.2 nm-thick Ir film ($\delta_{ave} = 3.2$ nm and $\delta_{\max} = 5$ nm) is not enough to generate a sufficient XRD signal. So these fibers are coated with 10 layers of 3.2 nm-thick Ir films and scanned for XRD again. The result is shown in Figure 6(a). The peak appears at 40.8 °, which indicates that the film is composed of crystals. The crystalline size is estimated to be about 8 nm.

Additionally, after XRD characterization, the same sample is studied by TEM (a JEOL 1200EX TEM with a 1.4 Å resolution). For the TEM sample preparation, a liquid resin is used with plasticizers and then mixed together with milkweed fibers. They are put into a vacuum chamber to drive air out of the liquid and the liquid flows into the hollow part of the fibers. This liquid mixture is poured in a mold and allowed to slowly polymerize at room temperature. After the solidification, this resin with fibers is sliced into thin pieces as the samples for TEM study. The low-magnified TEM images of 10 layers of 3.2 nm-thick Ir films coated on milkweed fiber is shown in Figure 6(b). We can see the maximum film thickness appears at the top and the thickness decreases gradually. Figure 6(c) shows the diffraction pattern of 10 layers of 3.2 nm-thick Ir films. The bright spots in the diffraction pattern show the existence of nanocrystals clearly. The high-resolution TEM image is shown in Figure 6(d). The yellow parallel lines show the lattice orientation. The different orientations of the lattice confirm the nanocrystalline structure of the Ir films on milkweed fiber.

## D. Physical mechanism behind the observed temperature-dependent behavior of thermal conductivity

For the characteristic size we revealed using the residual unified thermal resistivity, it includes the effect of point defect scattering, surface scattering, and grain boundary scattering as: $\tau_0^{-1} = \tau_{defect}^{-1} + \tau_{grain}^{-1} + \tau_{surface}^{-1}$ . The characterization length groups all the effects of point defect scattering, grain boundary scattering and surface scattering as $l_0^{-1} = \left(v_F \tau_{defect}\right)^{-1} + \left(v_F \tau_{grain}\right)^{-1} + \left(v_F \tau_{surface}\right)^{-1}$. In our previous work, the weak dependence of thermal conductivity on Ir film thickness proved that the surface scattering has little effect. Rather, the grain boundary scattering plays the major role in scattering electrons. Therefore, to first order estimation, the thermal resistance relation can be written as $l_{grain}/\kappa = l_{grain}/\kappa_c + R$. Here, $\kappa_c$ is the thermal conductivity of the bulk Ir and $R$ is the interfacial thermal resistance. Under this scenario, we can calculate the interface thermal conductance as $G = R^{-1} = (l_{grain}/\kappa - l_{grain}/\kappa_c)^{-1}$. The results are shown in Figure 7 and compared with the Al/Cu interface thermal conductance. The calculated Ir/Ir thermal conductance is much larger than that of the Al/Cu interface. This is because the Al/Cu interface is more highly mismatched than the Ir/Ir interface.

The electron's specific heat is proportional to $T$ when $T$ is not too high ($=\gamma T$). The observed thermal conductance variation with temperature is mostly determined by the specific heat of electrons. To check this point, *G/T* is also calculated and shown in the inset of Figure 7. *G/T* in fact represents a unified interface thermal conductance, and gives more direct information about the electron scattering behavior at the grain boundaries. Consequently, a unified interface thermal resistance: *RT* can also be used for studying the electron scattering behavior at the grain boundary. From the inset in Figure 7, we can see that *G/T* shows very weak temperature dependence. Its value changes from $2.61\times10^7$ W/m$^2$K$^2$ at room temperature to $2.27\times10^7$ W/m$^2$K$^2$ at 43 K. This indicates that interfacial thermal conductance is proportional to temperature and this temperature factor stems from the electron heat capacity. If point defect scattering is not considered and surface scattering is specular,[7,8] this interface thermal resistance is induced by the fact that some electrons are reflected instead of transmitting through the grain boundaries. Some reflected electrons could exchange energy with phonons at the grain boundaries before they are reflected back. Then these phonons exchange energy with phonons on the other side of the grain boundaries. In this case, the reflected electrons still have some of their energy transmitted across the grain boundaries. According to Mayadas-Shatzkes (MS) model,[27,28] the effective electron reflection coefficient by the grain boundary can be obtained and shown in Figure 7. As we can see, the electron reflection coefficient is large and almost constant. The electron reflection coefficient is 87.2% at room temperature. This means most of the electrons which scatter with the grain boundary are reflected back. This value becomes 88.7% when temperature goes down to 43 K. The very weak temperature dependent reflection coefficient indicates that the chance of electrons transport through grain boundaries is almost temperature-independent. The slightly higher grain boundary reflection coefficient at low temperatures gives rise to the slightly lower unified interface thermal conductance as indicated in the inset. It is noted that the grain boundary electron reflection coefficient we report here includes the effect of electron-phonon energy exchange adjacent to grain boundaries, and the phonon-phonon energy exchange across grain boundaries. Therefore, the real electron reflection coefficient should be a little higher than the values reported in Figure 7.

## V. UNCERTAINTY ANALYSIS

The relative error of length and diameter measurement with SEM, and electrical current and electrical resistance measurement are estimated as 1% and 0.5% respectively. The relative error of the Ir film thickness measurement is 2% determined by the quartz crystal microbalance in the sputtering system. For thermal diffusivity, every value is measured twenty times and the average value is determined as the final result. The maximum relative error for the fitting process is 10% but the real error is much smaller than 10%. Through fitting the thermal diffusivity difference ($\Delta\alpha_{eff}$), the average absolute error is $1.35\times10^{-8}$ $m^2/s$. The relative error of $\Delta\alpha_{eff}$ is then 6%. The fitting of volumetric specific heat shows a relative error of 6.4% and the volumetric specific heat is measured four times. So the relative error of average volumetric specific heat is 3.2%. Finally the relative error of the thermal conductivity of the Ir film is estimated as 7.2%.

## VI. CONCLUSION

In this work, the thermal conductivity of bio-supported average 3.2 nm-thin Ir film was characterized for the first time from room temperature down to 43 K. Close to two orders of magnitude reduction was observed for $\kappa$ of the film at low temperatures. $\kappa$ of the film increased with increasing temperature while that of bulk Ir decreased against temperature. We introduced a unified thermal resistivity (Θ) to explain the completely different $\kappa\sim T$ relation of the 3.2 nm film and the bulk Ir. It was found that the 3.2 nm film and the bulk Ir share the similar trend for $\Theta\sim T$ relation. At 0 K limit, the bulk Ir has a zero residual Θ while the 3.2 nm film has a very large residual Θ (5.5 $mK^2/W$), which dominated the overall unified thermal resistivity. The unified thermal resistivity played a critical role in quantitatively explaining the effect of defect in scattering electron during heat conduction. The evaluated interfacial thermal conductance among the grain boundaries was larger than that of the Al/Cu interface. It was proportional to temperature, and this relation was confirmed by the weak temperature dependent unified interfacial thermal conductance. It was found that the electron reflection coefficient was large (88%) and almost temperature independent.

## ACKNOWLEDGEMENT

Support of this work by Army Research Office (W911NF-12-1-0272), Office of Naval Research (N000141210603), and National Science Foundation (CBET1235852, CMMI1264399) is gratefully acknowledged. X.W thanks the partial support of the "Eastern Scholar" Program of Shanghai, China. We thank Christopher Reilly for careful proofreading of the manuscript.

**LIST OF TABLES AND FIGURES**

FIG. 1 (a) A milkweed seed and floss. (b) SEM image of a single milkweed fiber suspended across two electrodes (the long sample). The inset shows the floss surface. (c) SEM image of the milkweed fiber cross section. (d) Profile of the milkweed fiber cross section coated with a layer of Ir, and the definition of maximum thickness $\delta_{\max}$. The average thickness of the Ir film is $\delta_{ave} = 2\delta_{\max}/\pi$.

FIG. 2 (a) Schematic of the experimental principle of the TET technique to characterize the thermal diffusivity of the sample. (b) SEM image of a coated milkweed fiber connected across two electrodes (the short sample). (c) A typical *V*-*t* profile recorded by the oscilloscope for the sample shown in Figure (b) induced by the step DC current. The result is for the sample coated with the first Ir layer ($\delta_{1,\max}$=15 nm). (d) TET fitting results for the sample at room temperature. The figure consists of the normalized experimental temperature rise, theoretical fitting results, and other two fitting curves with ±10% variation of $\alpha_{eff}$ to demonstrate the uncertainty of the fitting process.

FIG. 3 Measured effective thermal diffusivity of the milkweed floss coated with different layers of Ir films. "$\Delta\alpha_{eff}$" is the effective thermal diffusivity difference between the 19.2 nm film ("15+5+5+5 nm" case whose $\delta_{ave}$=19.2 nm) and 9.6 nm film ("15 nm" case whose $\delta_{ave}$=9.6). "$\Delta\alpha_{eff}$ linear fit" represents linear fitting of "$\Delta\alpha_{eff}$" variation against temperature. "$\Delta\alpha_{eff,1}$" is the effective thermal diffusivity increment induced by each 3.2 nm-thick Ir layer. The solid curves are to show the trends of effective thermal diffusivity changing with temperature. The inset shows the thermal diffusivity changes against the number of film layers linearly to demonstrate that each 3.2 nm-thick Ir film

indeed has the same thermal conductivity, and follows the theory expressed by Eq. (1).

FIG. 4 Temperature dependent volumetric specific heat of milkweed and microcrystalline cellulose. The inset in the upper left corner shows the effective and intrinsic thermal conductivity of the milkweed fiber. The inset in the bottom right corner shows the effective thermal conductivity of the Ir-coated milkweed fiber.

FIG. 5 Temperature dependence of unified thermal resistivity of the 3.2 nm-thick Ir film and the bulk Ir (for comparison). "3.2 nm" is the data calculated from the linearly fitted $\Delta\alpha_{eff}$ shown in Figure 3. "Imperfection" represents $\Theta_{imper}$ induced by the imperfect structure in the film. The left inset shows the thermal conductivity variation against temperature and the right inset shows the orders of magnitude reduction of film's thermal conductivity from that of the bulk Ir (data from White, *et al.*). In the left inset, the "3.2 nm" depicts the thermal conductivity obtained directly from $\Delta\alpha_{eff}$ while the "3.2 nm_fit" shows the thermal conductivity obtained from the linear fitting values of $\Delta\alpha_{eff}$ (shown in Figure 3).

FIG. 6 (a) XRD pattern of 10 layers of 3.2 nm-thick Ir films on milkweed fibers. The peak appears at 40.8°, which indicates that the Ir film is composed of crystals. The crystalline size is estimated at about 8 nm. (b) Low-magnified TEM image of 10 layers of 3.2 nm-thick Ir films coated on milkweed fibers. (c) The diffraction pattern of 10 layers of 3.2 nm-thick Ir films. The bright spots in the diffraction pattern show the existence of nanocrystals clearly. (d) High-resolution TEM picture of the Ir film. The yellow parallel lines show the lattice orientation.

FIG. 7 Temperature dependent interfacial thermal conductance and electron reflection coefficient. "*G* of Ir/Ir" is the results of this work. For comparison, "*G* of Al/Cu (exp)" is the experimental results of Al/Cu interfacial thermal conductance and "*G* of Al/Cu (DMM)" is the prediction values of Al/Cu interfacial thermal conductance according to the diffusive mismatch model (DMM).[29] The inset shows the variation of *G/T* against temperature to demonstrate that the *G*-*T* relation shown in the figure mainly comes from the electron's specific heat against *T*.

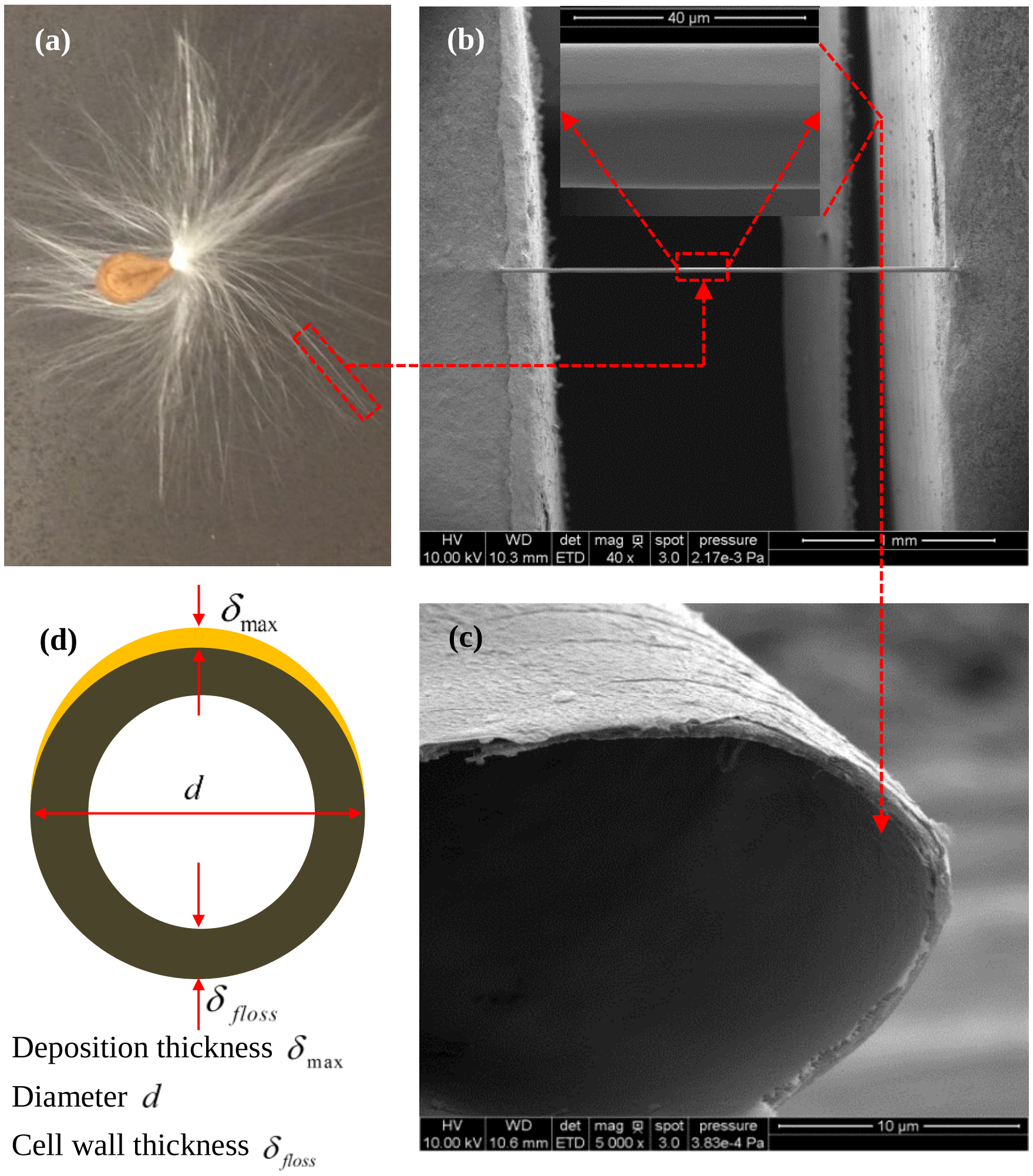

FIG. 1. (a) A milkweed seed and floss. (b) SEM image of a single milkweed fiber suspended across two electrodes (the long sample). The inset shows the floss surface. (c) SEM image of the milkweed fiber cross section. (d) Profile of the milkweed fiber cross section coated with a layer of Ir, and the definition of maximum thickness $\delta_{\max}$. The average thickness of the Ir film is $\delta_{ave} = 2\delta_{\max}/\pi$.

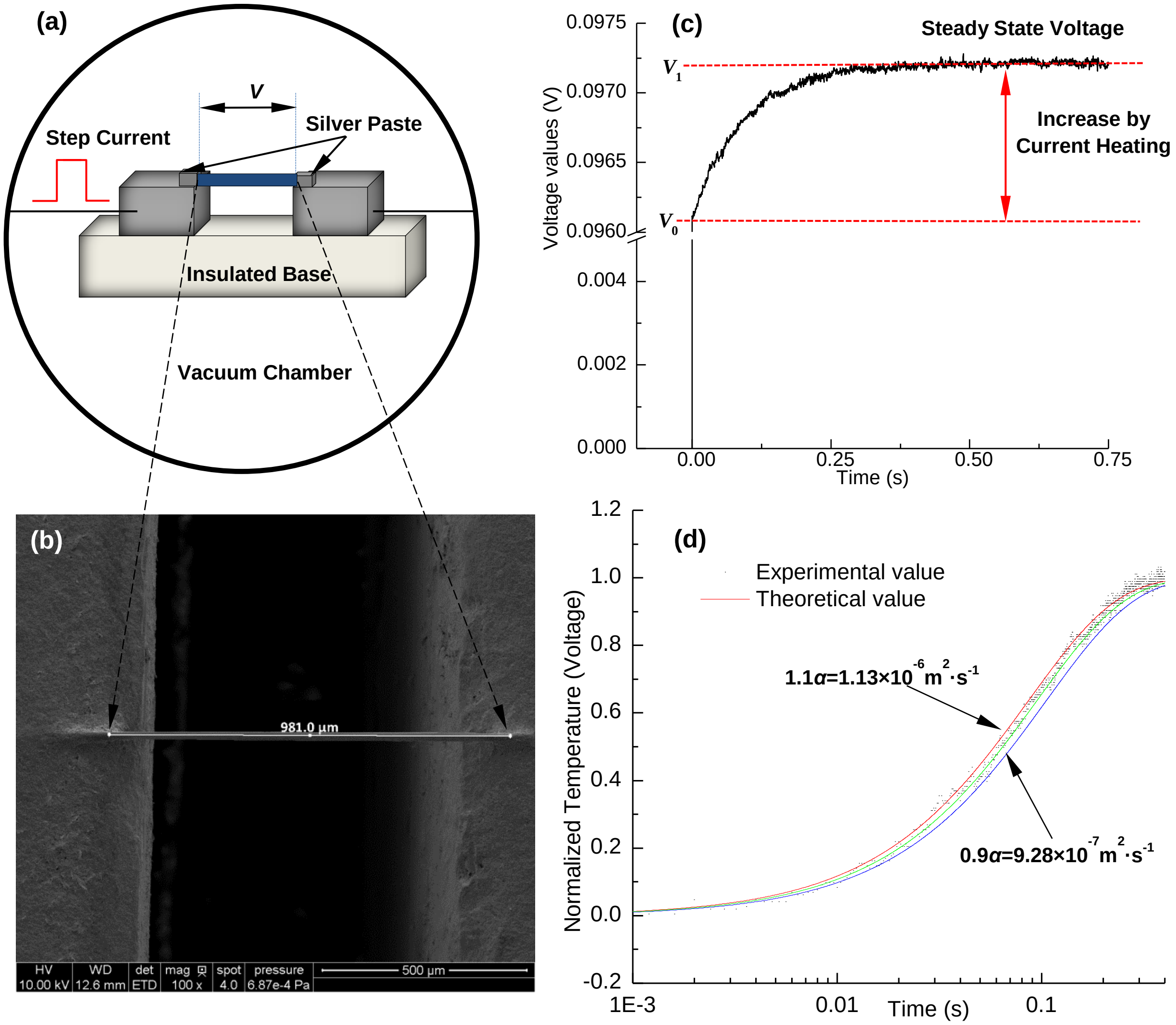

FIG. 2. (a) Schematic of the experimental principle of the TET technique to characterize the thermal diffusivity of the sample. (b) SEM image of a coated milkweed fiber connected across two electrodes (the short sample). (c) A typical *V*-*t* profile recorded by the oscilloscope for the sample shown in Figure (b) induced by the step DC current. The result is for the sample coated with the first Ir layer ($\delta_{1,\max}$ =15 nm). (d) TET fitting results for the sample at room temperature. The figure consists of the normalized experimental temperature rise, theoretical fitting results, and other two fitting curves with ±10% variation of $\alpha_{eff}$ to demonstrate the variation of the experimental data around the best theoretical fitting.

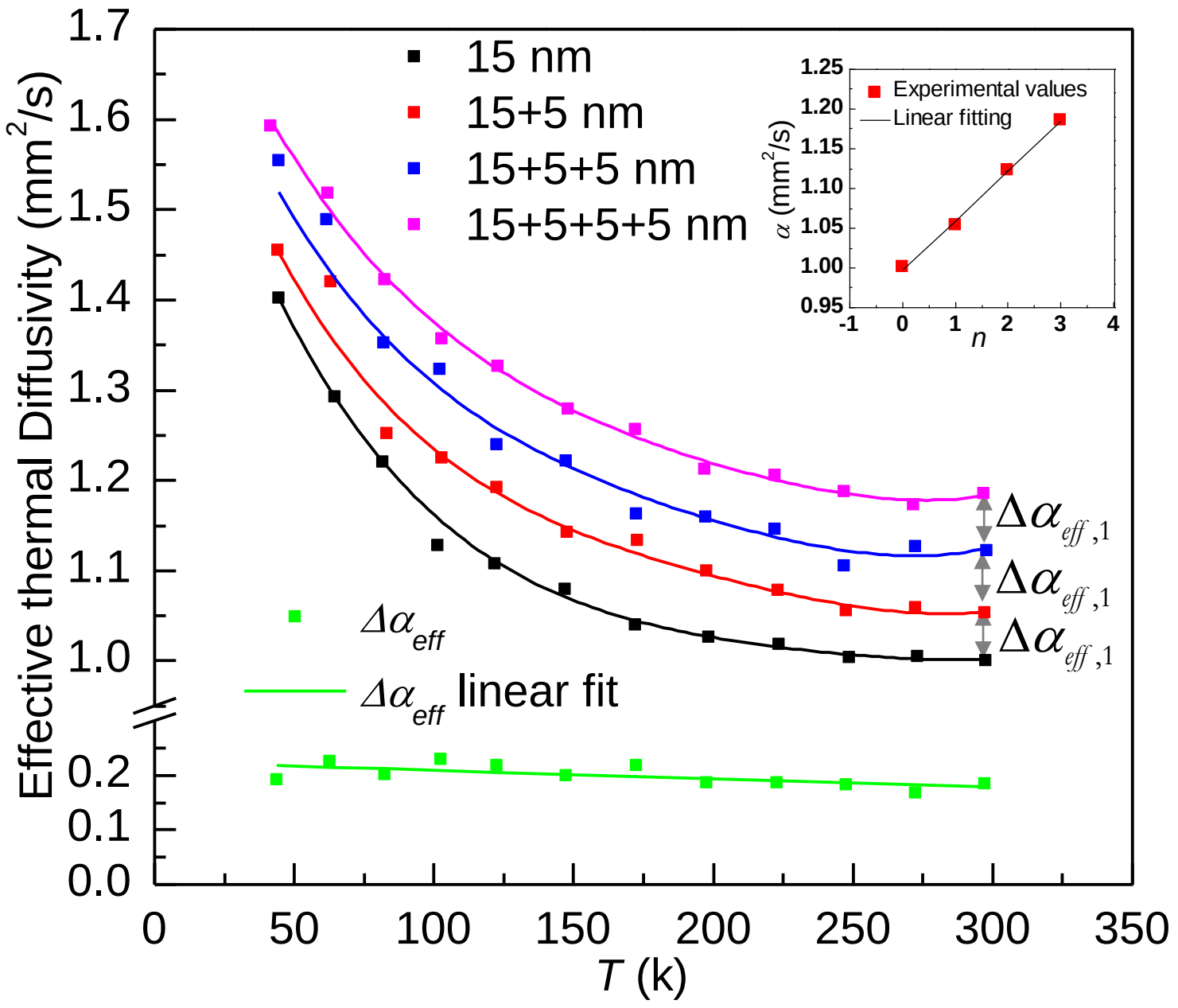

FIG. 3. Measured effective thermal diffusivity of the milkweed floss coated with different layers of Ir films. "$\Delta\alpha_{eff}$" is the effective thermal diffusivity difference between the 19.2 nm film ("15+5+5+5 nm" case whose $\delta_{\mathrm{ave}}$=19.2 nm) and 9.6 nm film ("15 nm" case whose $\delta_{\mathrm{ave}}$=9.6). "$\Delta\alpha_{eff}$ linear fit" represents linear fitting of "$\Delta\alpha_{eff}$" variation against temperature. "$\Delta\alpha_{eff,1}$" is the effective thermal diffusivity increment induced by each 3.2 nm-thick Ir layer. The solid curves are to show the trends of effective thermal diffusivity changing with temperature. The inset shows the thermal diffusivity changes against the number of film layers linearly to demonstrate that each 3.2 nm-thick Ir film indeed has the same thermal conductivity, and follows the theory expressed by Eq. (1).

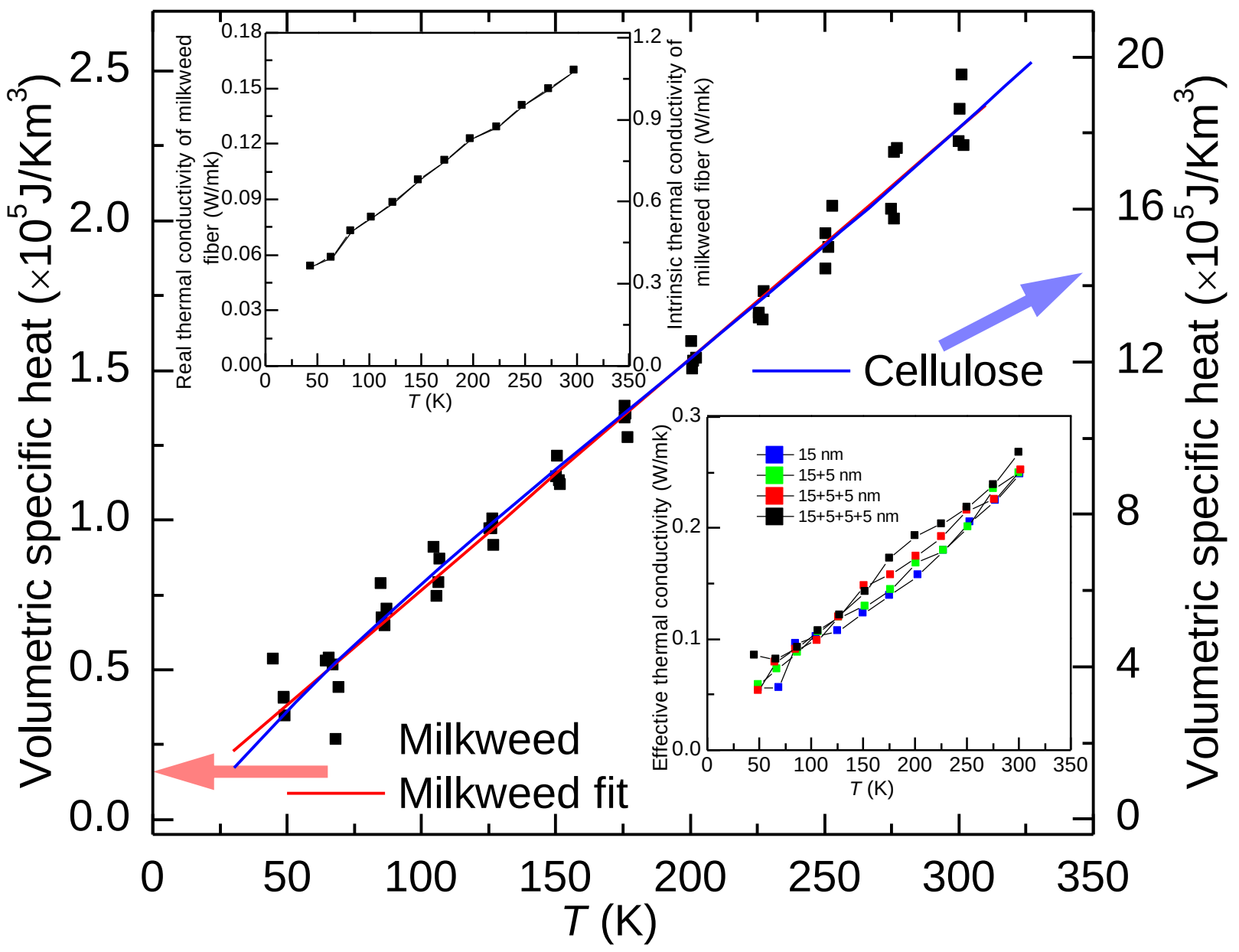

FIG. 4. Temperature dependent volumetric specific heat of milkweed and microcrystalline cellulose. The inset in the upper left corner shows the effective and intrinsic thermal conductivity of the milkweed fiber. The inset in the bottom right corner shows the effective thermal conductivity of the Ir-coated milkweed fiber.

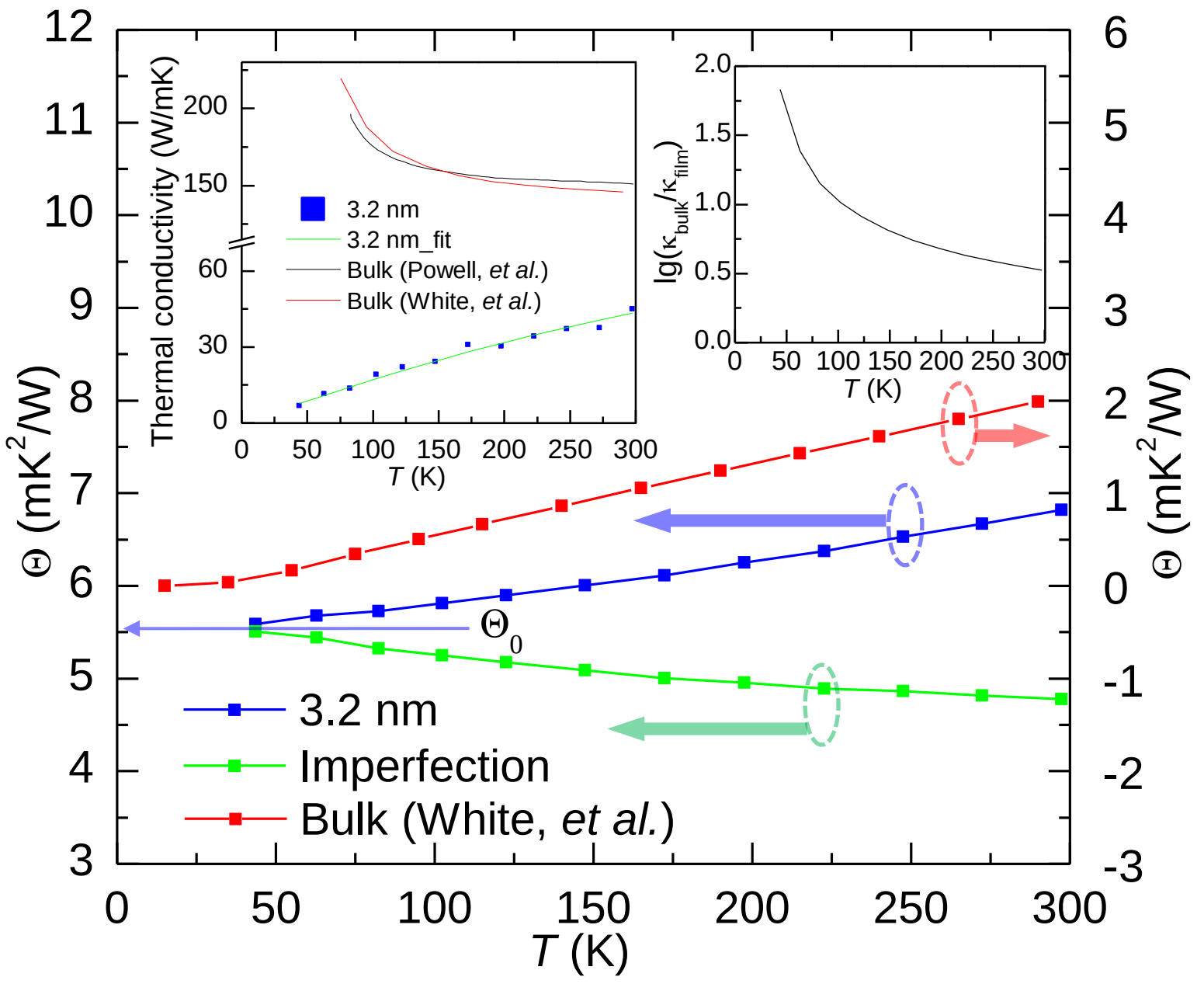

FIG. 5. Temperature dependence of unified thermal resistivity of the 3.2 nm-thick Ir film and the bulk Ir (for comparison). "3.2 nm" is the data calculated from the linearly fitted $\Delta\alpha_{eff}$ shown in Figure 3. "Imperfection" represents $\Theta_{imper}$ induced by the imperfect structure in the film. The left inset shows the thermal conductivity variation against temperature and the right inset shows the orders of magnitude reduction of film's thermal conductivity from that of bulk Ir (data from White, *et al.*). In the left inset, the "3.2 nm" depicts the thermal conductivity obtained directly from $\Delta\alpha_{eff}$ while the "3.2 nm_fit" shows the thermal conductivity obtained from the linear fitting values of $\Delta\alpha_{eff}$ (shown in Figure 3).

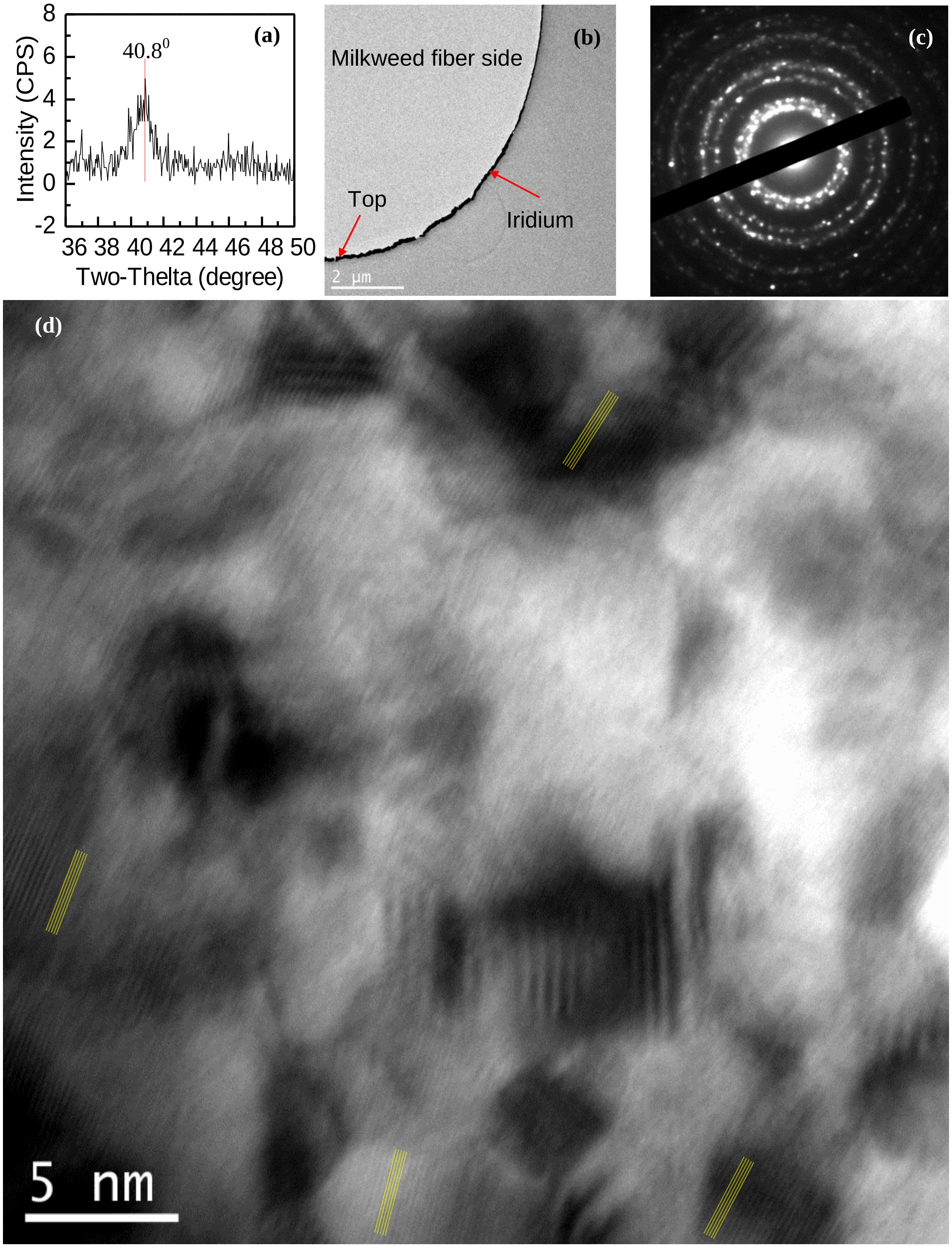

FIG. 6. (a) XRD pattern of 10 layers of 3.2 nm-thick Ir films on milkweed fibers. The peak appears at 40.8°, which indicates that the Ir film is composed of crystals. The crystalline size is estimated at about 8 nm. (b) Low-magnified TEM image of 10 layers of 3.2 nm-thick Ir films coated on milkweed fibers. (c) The diffraction pattern of 10 layers of 3.2 nm-thick Ir films. The bright spots in the diffraction pattern show the existence of nanocrystals clearly. (d) High-resolution TEM picture of the Ir film. The yellow parallel lines show the lattice orientation.

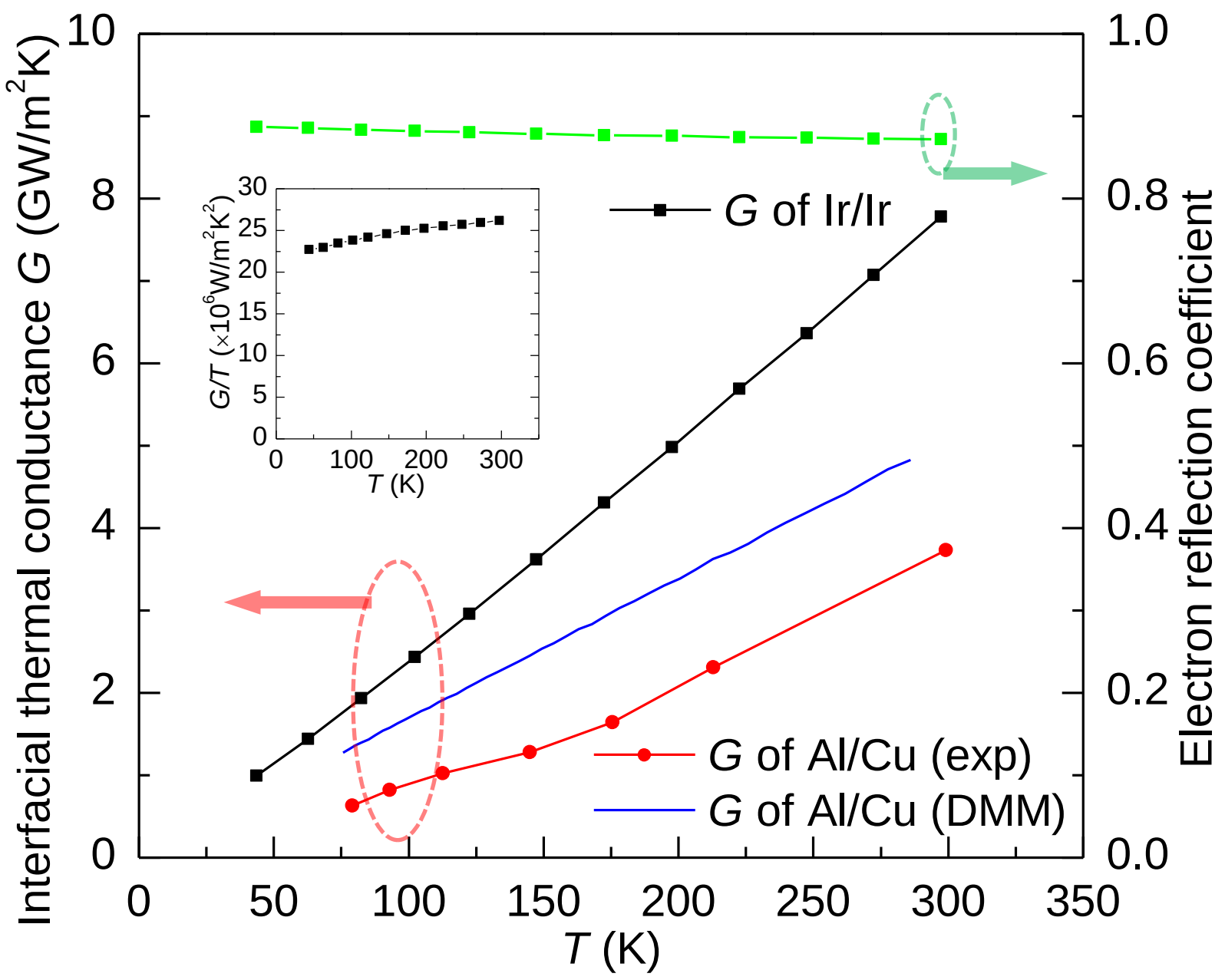

FIG. 7. Temperature dependent interfacial thermal conductance and electron reflection coefficient. "*G* of Ir/Ir" is the results of this work. For comparison, "*G* of Al/Cu (exp)" is the experimental results of Al/Cu interfacial thermal conductance and "*G* of Al/Cu (DMM)" is the prediction values of Al/Cu interfacial thermal conductance according to the diffusive mismatch model (DMM).[29] The inset shows the variation of *G/T* against temperature to demonstrate that the *G*-*T* relation shown in the figure mainly comes from the electron's specific heat change against *T*.

# Table of Contents Graphics

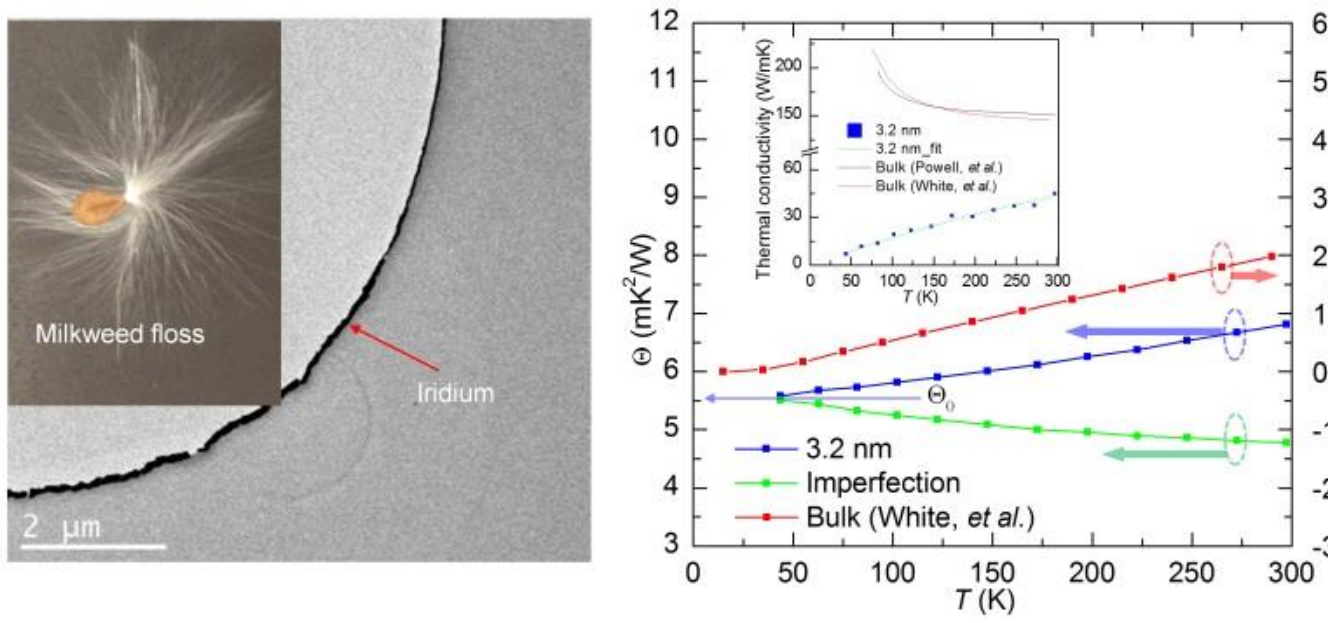